\documentstyle[prd,aps,preprint,tighten,epsfig]{revtex}

\begin{document}

\draft
\preprint{\begin{tabular}{r}
{\bf hep-ph/0006112} \\
LUM-00-06 \\
May 2000
\end{tabular}}

\title{Can CPT Symmetry Be Tested With $K^0$ vs $\bar{K}^0
\rightarrow \pi^+\pi^-\pi^0$ Decays?} 
\author{\bf Zhi-zhong Xing}
\address{Sektion Physik, Universit$\it\ddot{a}$t M$\it\ddot{u}$nchen,
Theresienstrasse 37A, 80333 M$\it\ddot{u}$nchen, Germany \\
{\it Electronic address: xing@theorie.physik.uni-muenchen.de} }
\maketitle

\begin{abstract}
We show that the CP-violating effect in 
$K^0$ vs $\bar K^0 \rightarrow \pi^+\pi^-\pi^0$ decays
differs from that in $K_{\rm L} \rightarrow \pi^+\pi^-$, 
$K_{\rm L} \rightarrow \pi^0\pi^0$ or the semileptonic $K_{\rm L}$ 
transitions, if there exists CPT violation in $K^0$-$\bar{K}^0$ mixing. 
A delicate measurement of this difference in the KTeV experiment 
and at the $\phi$ factory will provide a new test of CPT symmetry 
in the neutral kaon system.
\end{abstract}

\pacs{PACS number(s): 11.30. Er, 13.25. Es, 14.40. Aq} 

\newpage

The $K^0$-$\bar{K}^0$ mixing system has been playing an important role 
in particle physics for testing fundamental symmetries (such as CP, T 
and CPT) and examining conservation laws (such as $\rm\Delta S = \Delta Q$). 
The existing experimental evidence for CPT invariance in the mixing and
decays of neutral kaon mesons remains rather poor \cite{PDG98}: 
it is not excluded that the strength of CPT-violating interactions could 
be as large as about ten percentage of that of CP-violating interactions. 
This unsatisfactory situation will be improved in the near future, in
particular after a variety of more delicate measurements are carried 
out in the KTeV experiment \cite{KTeV} and at the $\phi$ factory \cite{Phi}. 

There are several possibilities to pin down CPT violation in
$K^0$-$\bar{K}^0$ mixing with the decays of $K_{\rm S}$ and $K_{\rm L}$
mesons into the two-pion and (or) the semileptonic states \cite{Thomson}.
Recently a new possibility to test CPT symmetry, with the help of
neutral kaon decays into the three-pion states, has been
pointed out in Ref. \cite{Sanda}. The idea is simply that
the CP-violating effect induced by $K^0$-$\bar K^0$ mixing in 
$K^0$ vs $\bar K^0 \rightarrow \pi^+\pi^-\pi^0$ transitions should not be
identical to that in $K_{\rm L}\rightarrow \pi^+\pi^-$, 
$K_{\rm L} \rightarrow \pi^0\pi^0$ or the 
semileptonic $K_{\rm L}$ decays, if CPT symmetry is also violated. Thus a
careful comparison between these two types of CP-violating effects
may provide a robust test of CPT invariance in $K^0$-$\bar K^0$ mixing. 

This paper aims at elaborating the preliminary idea proposed 
in Ref. \cite{Sanda} and presenting the time-dependent CP asymmetry
between $K^0(t) \rightarrow \pi^+\pi^-\pi^0$ and
$\bar{K}^0(t) \rightarrow \pi^+\pi^-\pi^0$ decays. 
Our result is expected to be useful for the forth-coming experiments
of kaon physics. 

Let us explain the main idea concisely. The mass eigenstates of
$K^0$ and $\bar K^0$ mesons can in general be written as 
\begin{eqnarray}
|K_{\rm S} \rangle & = & \frac{1}{\sqrt{|p^{~}_1|^2 + |q^{~}_1|^2}} 
\left ( p^{~}_1 |K^0\rangle + q^{~}_1 |\bar{K}^0 \rangle \right ) \; ,
\nonumber \\
|K_{\rm L} \rangle & = & \frac{1}{\sqrt{|p^{~}_2|^2 + |q^{~}_2|^2}} 
\left ( p^{~}_2 |K^0\rangle - q^{~}_2 |\bar{K}^0\rangle \right ) \; ,
%               (1)
\end{eqnarray}
where $p^{~}_i$ and $q^{~}_i$ (for $i=1,2$) are complex mixing parameters. Note that
$p^{~}_1 =p^{~}_2$ and $q^{~}_1 = q^{~}_2$ follow from CPT symmetry. 
The traditional characteristic measurables of CP violation in the
$K^0$-$\bar{K}^0$ system \cite{PDG98},
$\eta_{+-}$, $\eta_{00}$ and $\delta$, are all related to the 
$(p^{~}_2, q^{~}_2)$ parameters. For example,
\begin{equation}
\delta \; =\; \frac{|p^{~}_2|^2 - |q^{~}_2|^2}{|p^{~}_2|^2 + |q^{~}_2|^2} \; 
%               (2)
\end{equation}
in the absence of $\rm\Delta S = - \Delta Q$ interactions.
A measurement of CP violation associated with
\begin{equation}
\kappa \; = \; \frac{|p^{~}_1|^2 - |q^{~}_1|^2}{|p^{~}_1|^2 + |q^{~}_1|^2} \; 
%               (3)
\end{equation}
has been assumed to be extremely difficult, if not impossible, due to
the rapid decay of the $K_{\rm S}$ meson to the two-pion or the semileptonic
state. Nevertheless, we shall show that $\kappa$ can be 
measured from the rate asymmetry of $K^0$ and $\bar{K}^0$ mesons
decaying into the three-pion state $\pi^+\pi^-\pi^0$. The difference
between $\delta$ and $\kappa$ signifies CPT violation in $K^0$-$\bar{K}^0$
mixing.

The CP eigenvalue for the $\pi^+\pi^-\pi^0$ final state is given by
$(-1)^{l+1}$, where $l$ is the relative angular momentum between
$\pi^+$ and $\pi^-$. Since the sum of the masses of three pions is
close to the kaon mass, the pions have quite low kinetic energy
$E_{\rm CM}(\pi)$ in the kaon rest-frame, and the states with $l>0$ 
are suppressed by the centrifugal barrier \cite{CPLEAR}. Therefore the 
$K_{\rm L}$ meson decays dominantly into
the kinematics-favored ($l=0$) and CP-allowed (CP = $-$1) 
$\pi^+\pi^-\pi^0$ state. The decay amplitude of $K_{\rm S}\rightarrow
\pi^+\pi^-\pi^0$ consists of both the kinematics-suppressed ($l=1$) 
but CP-allowed (CP=+1) component, and the kinematics-favored
($l=0$) but CP-forbidden (CP=$-$1) component. This implies an
interesting Dalitz-plot distribution for the $K_{\rm S}\rightarrow
\pi^+\pi^-\pi^0$ transition: it is symmetric with respect to 
$\pi^+$ and $\pi^-$ for the CP-violating amplitude, but anti-symmetric 
for the CP-conserving amplitude.
Let the ratio of $K_{\rm S}$ and $K_{\rm L}$ decay amplitudes be
\begin{equation}
\eta^{~}_{+-0} \; = \; \frac{A (K_{\rm S} \rightarrow \pi^+\pi^-\pi^0)}
{A(K_{\rm L} \rightarrow \pi^+\pi^-\pi^0)} \; .
%               (4)
\end{equation}
It is clear that $\eta^{~}_{+-0}$ depends only upon the
CP-violating component of $A( K_{\rm S} \rightarrow \pi^+\pi^-\pi^0)$, if
data are integrated over the whole Dalitz plot \cite{CPLEAR,Nakada98}. 
The time-dependent
rates for the initially pure $K^0$ and $\bar K^0$ states decaying into
$\pi^+\pi^-\pi^0$, denoted by ${\cal R}(t)$ and $\bar{\cal R}(t)$ 
respectively, can be calculated with the help of Eqs. (1)
and (4). We arrive at 
\begin{eqnarray}
{\cal R}(t) & \propto & \left [ |q^{~}_1|^2 + |q^{~}_2|^2 |\eta_{+-0}|^2 
e^{-\Delta \Gamma t} + 2 {\rm Re} \left (q^*_1 q^{~}_2 \eta_{+-0}
e^{i \Delta m t} \right ) e^{-\Delta \Gamma t/2} \right ] \; ,
\nonumber \\
\bar{\cal R}(t) & \propto & \left [ |p^{~}_1|^2 + |p^{~}_2|^2 |\eta_{+-0}|^2 
e^{-\Delta \Gamma t} - 2 {\rm Re} \left (p^*_1 p^{~}_2 \eta_{+-0}
e^{i \Delta m t} \right ) e^{-\Delta \Gamma t/2} \right ] \; ,
%               (5)
\end{eqnarray}
where $\Delta m >0$ and $\Delta \Gamma >0$ are the mass difference 
and the width difference of $K_{\rm S}$ and $K_{\rm L}$ mesons, respectively.
To a good degree of accuracy, we obtain the following CP asymmetry:
\begin{eqnarray}
{\cal A}(t) & = & \frac{\bar{\cal R}(t) ~ - ~ {\cal R}(t)}
{\bar{\cal R}(t) ~ + ~ {\cal R}(t)} \; \nonumber \\
\nonumber \\
& = & \kappa ~ - ~ 2 e^{-\Delta \Gamma t/2} \left [ {\rm Re}\eta^{~}_{+-0} 
\cos (\Delta m t) ~ - ~ {\rm Im}\eta^{~}_{+-0} \sin
(\Delta m t) \right ] \xi \; \nonumber \\
&  & ~~  ~ - ~ 2 e^{-\Delta \Gamma t/2} \left [ {\rm Re}\eta^{~}_{+-0} 
\sin (\Delta m t) ~ + ~ {\rm Im}\eta^{~}_{+-0} \cos
(\Delta m t) \right ] \zeta \;\; ,
%               (6)
\end{eqnarray}
in which
\begin{eqnarray}
\xi & = & \frac{{\rm Re} ( p^{~}_1 p^*_2 + q^{~}_1 q^*_2
)}{|p^{~}_1|^2 + |q^{~}_1|^2} \; \; , \nonumber \\
\zeta & = & \frac{{\rm Im} ( p^{~}_1 p^*_2 + q^{~}_1 q^*_2
)}{|p^{~}_1|^2 + |q^{~}_1|^2} \;\; .
%               (7)
\end{eqnarray}
Obviously $\kappa$ can be determined through
the measurement of ${\cal A}(t)$. In particular, the relationship
$\displaystyle\lim_{t\rightarrow \infty} {\cal A}(t) = \kappa$ holds
%%%%%%%%%%%%%%%%%%%%%%%%%%%
\footnote{Note that the result for 
$\displaystyle\lim_{t\rightarrow \infty} {\cal A}(t)$
given in Ref. \cite{Sanda} is not correct.}.
%%%%%%%%%%%%%%%%%%%%%%%%%%%

As we have emphasized, the difference between $\kappa$ and $\delta$ 
hints at CPT violation in $K^0$-$\bar{K}^0$ mixing. 
This point can be seen more clearly, if one adopts the popular
$(\epsilon, \Delta)$ parameters to describe CP- and CPT-violating
effects in the $K^0$-$\bar K^0$ mixing system:
\begin{eqnarray}
p^{~}_1 & = & 1 + \epsilon + \Delta \; , \nonumber \\
p^{~}_2 & = & 1 + \epsilon - \Delta \; , \nonumber \\
q^{~}_1 & = & 1 - \epsilon - \Delta \; , \nonumber \\
q^{~}_2 & = & 1 - \epsilon + \Delta \; .
%               (8)
\end{eqnarray}
With this notation we find
\begin{eqnarray}
\delta & = & 2 \left ( {\rm Re} \epsilon ~ - ~ {\rm Re} \Delta \right
) \; , \nonumber \\
\kappa & = & 2 \left ( {\rm Re} \epsilon ~ + ~ {\rm Re} \Delta \right
) \; .
%               (9)
\end{eqnarray}
If $|{\rm Re}\Delta| / {\rm Re}\epsilon \sim 0.1$, then the difference
$\kappa - \delta = 4 {\rm Re}\Delta$ can be as large as $0.4 {\rm Re}
\epsilon \sim 6.6 \times 10^{-4}$ in magnitude, where the experimental
value ${\rm Re}\epsilon
\approx 1.65 \times 10^{-3}$ has been used \cite{PDG98}.
Since both $\epsilon$ and $\Delta$ are small quantities, it turns out
that $\xi \approx 1$ and $\zeta \approx 0$ are good approximations.
Eq. (6) is therefore simplified as
\begin{equation}
{\cal A}(t) \; = \;
\kappa ~ - ~ 2 e^{-\Delta \Gamma t/2} \left [ {\rm Re}\eta^{~}_{+-0} 
\cos (\Delta m t) ~ - ~ {\rm Im}\eta^{~}_{+-0} \sin (\Delta m t) \right ] \; .
%               (10)
\end{equation}
In the neglect of CPT violation, i.e., $\kappa = 2 {\rm Re}\epsilon$, 
Eq. (10) just reproduces the result obtained in Ref. \cite{CPLEAR}.
For illustration, we plot the behavior of ${\cal A}(t)$ in Fig. 1,
in which $\kappa = 3\times 10^{-3}$ and $|\eta_{+-0}| =
5\times 10^{-3}$ have been typically taken
%%%%%%%%%%%%%%%%%%%%%%%
\footnote{The CPLEAR constraint on the CP-violating parameter
$\eta_{+-0}$ is $|\eta_{+-0}| < 0.017$ at the $90\%$ confidence
level \cite{CPLEAR}.}.
%%%%%%%%%%%%%%%%%%%%%%%
We observe that ${\cal A}(t)$ approaches $\kappa$ for $t \geq 5
\tau^{~}_{\rm S}$ and reaches $\kappa$ if $t \geq 10 \tau^{~}_{\rm S}$, where
$\tau^{~}_{\rm S}$ is the mean lifetime of the $K_{\rm S}$ meson. 
This implies a certain feasibility
to determine $\kappa$ from the time-dependent CP asymmetry ${\cal A}(t)$.

In the above discussions we have taken an integration over the whole
Dalitz plot, such that $\eta_{+-0}$ solely contains the CP-violating 
part of $A(K_{\rm S}\rightarrow \pi^+\pi^-\pi^0)$. To look at the
CP-conserving component of $A(K_{\rm S}\rightarrow \pi^+\pi^-\pi^0)$, one
may study the phase-space regions $E_{\rm CM}(\pi^+) > E_{\rm
CM}(\pi^-)$ and $E_{\rm CM}(\pi^+) < E_{\rm CM}(\pi^-)$ separately
\cite {CPLEAR}. In this case the corresponding CP asymmetries
between $\bar{\cal R}(t)$ and ${\cal R}(t)$ take the same form
as ${\cal A}(t)$ in Eq. (6) or Eq. (10), but $\eta_{+-0}$ should be
replaced by $(\eta_{+-0} \pm \lambda)$, where $\lambda$ denotes the
CP-conserving contribution to the ratio of $K_{\rm S}$ and 
$K_{\rm L}$ decay amplitudes \cite{CPLEAR}. 
Certainly the CP-violating parameter
$\kappa$ can still be determined from measuring the time dependence of 
the relevant decay rate asymmetries.

An accurate measurement of $\kappa$ from $K^0$ vs $\bar K^0
\rightarrow \pi^+\pi^-\pi^0$ is feasible 
at the $\phi$ factory, where a huge amount of
$K^0\bar K^0$ events can be coherently produced \cite{Phi}. Choosing the
semileptonic decay of one kaon to tag the flavor of the other kaon
decaying into $\pi^+\pi^-\pi^0$ on the $\phi$ resonance, one should be 
able to construct the time-dependent rate asymmetry between 
$K^0(t) \rightarrow \pi^+\pi^-\pi^0$ and 
$\bar K^0(t) \rightarrow \pi^+\pi^-\pi^0$ decays in
a way similar to Eq. (6). It is also expected that the KTeV experiment
at Fermilab \cite{KTeV} may measure $\kappa$ and $\delta$ 
to a good degree of accuracy.

To conclude, the CP-violating effect induced by $K^0$-$\bar K^0$
mixing in $K^0$ vs $\bar K^0 \rightarrow \pi^+\pi^-\pi^0$ decays
is possible to deviate somehow from that in $K_{\rm L} \rightarrow
\pi\pi$ or the semileptonic $K_{\rm L}$ transitions due to the
violation of CPT symmetry. Measuring this difference may serve as a
new test of CPT invariance in the neutral kaon system. 

%The author is grateful to L.B. Okun, A.I. Sanda, and S.Y. Tsai for 
%useful discussions.

\newpage

%%%%%%%%%%%%%%%%%%%% Fig. 1 %%%%%%%%%%%%%%%%
\begin{figure}[t]
\epsfig{file=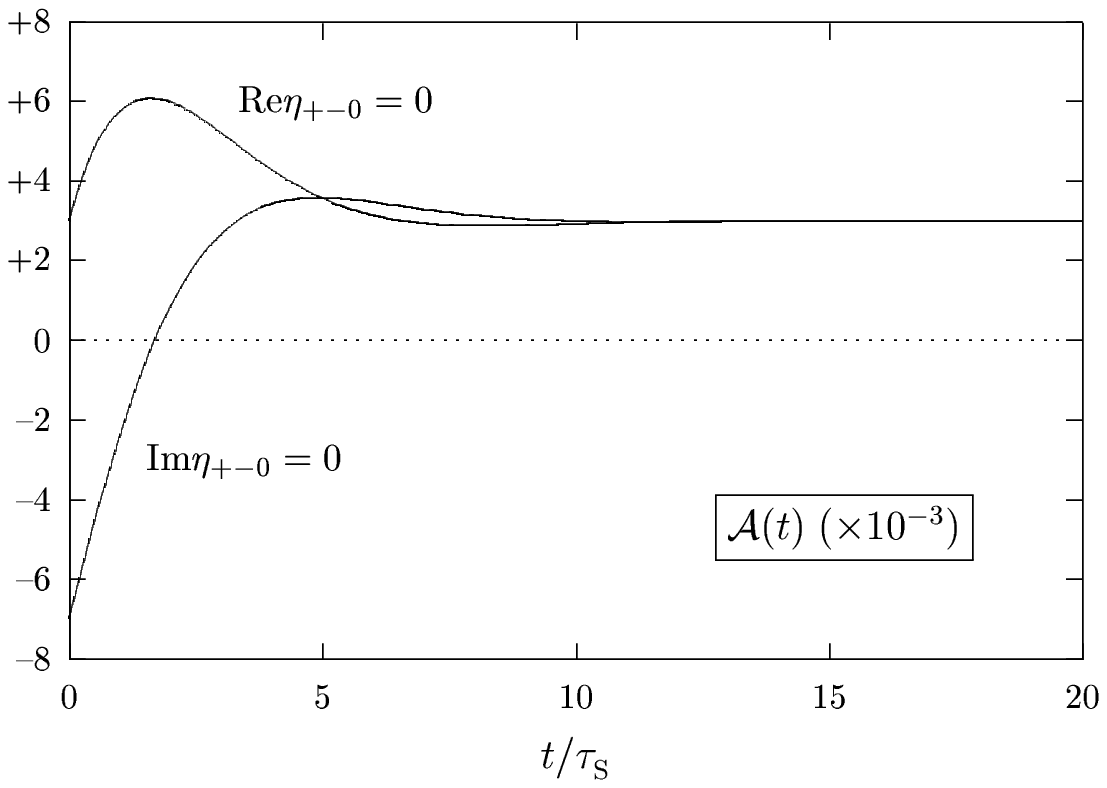,bbllx=3cm,bblly=10cm,bburx=20cm,bbury=29cm,%
width=15.5cm,height=20cm,angle=0,clip=}
\vspace{-9.1cm}
\caption{Illustrative plot for the CP asymmetry ${\cal A}(t)$
given in Eq. (10) with the typical inputs $\kappa = 3\times 10^{-3}$ and
$|\eta_{+-0}| = 5\times 10^{-3}$. Here $\tau^{~}_{\rm S}$ is the 
mean lifetime of the $K_{\rm S}$ meson.}
\end{figure}
%%%%%%%%%%%%%%%%%%%%%%%%%%%%%%%%%%%%%%%%%%%%

\end{document}